\documentclass{llncs}

\usepackage{graphicx}
\usepackage{amsmath}
\usepackage{url}

\begin{document}

\title{Spectral Properties of Adjacency and Distance Matrices for Various Networks}
\author{Krzysztof Malarz}
\institute{
  AGH University of Science and Technology,
  Faculty of Physics and Applied Computer Science,
  al. Mickiewicza 30, PL-30059 Krak\'ow, Poland.\\
  \email{malarz@agh.edu.pl}, 
  \url{http://home.agh.edu.pl/malarz/}
  }

\maketitle

\begin{abstract}
The spectral properties of the adjacency (connectivity) and distance matrix for various types of networks: 
exponential, scale-free (Albert--Barab\'asi) and classical random ones (Erd\H os--R\'enyi) are evaluated. 
The graph spectra for dense graph in the Erd\H os--R\'enyi model are derived analytically.
\end{abstract}


\section{Introduction}

Studies of the network structure seem to be essential for better understanding of many real-world complex systems \cite{Albert-rev,Dorogovtsev-rev,Newman-rev}.
Among these systems are social \cite{socio}, economic \cite{econo}, biological \cite{bio} systems or networks sensu stricto \cite{cmpsc} like Internet or World Wide Web. 
In the latter case effective algorithms for WWW content search are particularly desired. 
The {\tt Google} search engine of the network search bases on the eigenvector centrality \cite{Newman-rev,eigencentrality} 
which is well known in the social network analysis and not different from the Brin and Page algorithm \cite{Newman-rev,Brin}.
In this algorithm each vertex $i$ of the network is characterized by a positive weight $w_i$ proportional to the sum 
of the weights $\sum_j w_j$ of all vertexes which point to $i$, where $w_i$ are elements of the $i$-th eigenvector $\mathbf{w}$ 
of the graph adjacency matrix $\mathbf{A}$
\begin{equation}
\mathbf{A}\mathbf{w}=\lambda\mathbf{w}.
\end{equation}
The concept of eigenvector centrality allows distinguish between different importance of the links and thus is much richer than degree or node centrality \cite{Newman-2008}.
The adjacency matrix $\mathbf{A}$ of the network with $N$ nodes is square $N\times N$ large matrix which elements $a(i,j)$ shows number of (directed) links form node $i$ to $j$.
For undirected network this matrix is symmetrical.
For simple graphs (where no multiple edges are possible) this matrix is binary: $a(i,j)=1$ when nodes $i$--$j$ are linked together else $a(i,j)=0$.
The set of eigenvalues (or its density $\rho_A(\lambda)$) of the adjacency matrix $\mathbf{A}$ is called a graph/network spectrum.
The graph spectrum was examined \cite{Mehta,Cvetkovic} for classical random graphs (Erd{\H o}s--R\'enyi, ER) \cite{ER} and investigated numerically for scale-free networks \cite{AB} by Farkas et al. \cite{Farkas-2001,Farkas-2002}.
The spectra of complex networks were derived exactly for infinite random uncorrelated and correlated random tree-like graphs by Dorogovtsev et al. \cite{Dorogovtsev-2003}.
Several other examples of networks properties obtained by studies of graph spectra are given in Ref. \cite{net-eig-stu}.

	While many papers refer to eigenvalues of the adjacency matrices $\mathbf{A}$, less 
is known about the spectra of the distance matrices $\mathbf{D}$.
In the distance matrix $\mathbf{D}$ element $d(i,j)$ is the length of the shortest path between nodes $i$ and $j$.
On the other hand, whole branch of topological organic chemistry for alkens was 
developed for small graphs which symbolize alkens' structural formula \cite{Schultz}.
There, not only adjacency $\mathbf{A}$ and distance $\mathbf{D}$ matrix but also their 
sum $\mathbf{A}+\mathbf{D}$ spectral properties were investigated.

	The detailed description of the distance matrix construction during the network growth for the various 
network types is given in Ref. \cite{Malarz}.
Other solutions of this problem are also known; an example is the Floyd algorithm \cite{Floyd}.
During the network growth nodes may be attached to so far existing nodes randomly or according to some preferences $P$.
When this preference bases on nodes connectivity $k$, $P(k)\propto k$, the scale-free Albert--Barab\'asi (AB) \cite{AB} networks will appear. 
The pure random attachment ($P(k)=\text{const}$) leads to exponential nodes degree distribution.
New nodes may bring with itself one ($M=1$) or more ($M\ge2$) edges which serve as links to pre-existing graph.
For $M=1$ the tree-like structure appears, while for $M>1$ the cyclic path are available.
Let us recall that degree distributions $\pi(k)$ are $\pi(k)\propto k^{-\gamma}$, $\pi(k)\propto\exp(-k)$ and Poisson's one for AB, exponential and ER networks, respectively \cite{Albert-rev,Dorogovtsev-rev,Newman-rev}.

	Here we study numerically\footnote{with LAPACK procedure {\tt http://www.netlib.org/lapack/double/dsyev.f}} the graph spectra $\rho_A(\lambda)$ for growing networks with exponential degree distribution for $M=1$ and $M=2$.  
We check the eigenvalue density $\rho_D(\lambda)$ of the distance matrix $\mathbf{D}$ for AB, exponential and ER graphs.
In literature known to us these spectra was never checked before.

	The graph spectrum $\rho_A(\lambda)$ for dense graph in the ER model is derived analytically in Sec. \ref{sec-results-A} as well.
Here we profit much from Ref. \cite{Burda-2007}.

\section{Results and Discussion}
\label{sec-results}
Here we show densities of eigenvalues $\rho(\lambda)$ for matrices $\mathbf{A}$ and 
$\mathbf{D}$ for various kinds of networks.
Results are averaged over $N_{\text{run}}=10^2$ realizations of networks of $N=10^3$ nodes.

\subsection{Spectral Properties of Adjacency Matrix}
\label{sec-results-A}
\begin{figure}
\begin{center}
(a) \includegraphics[width=.6\textwidth]{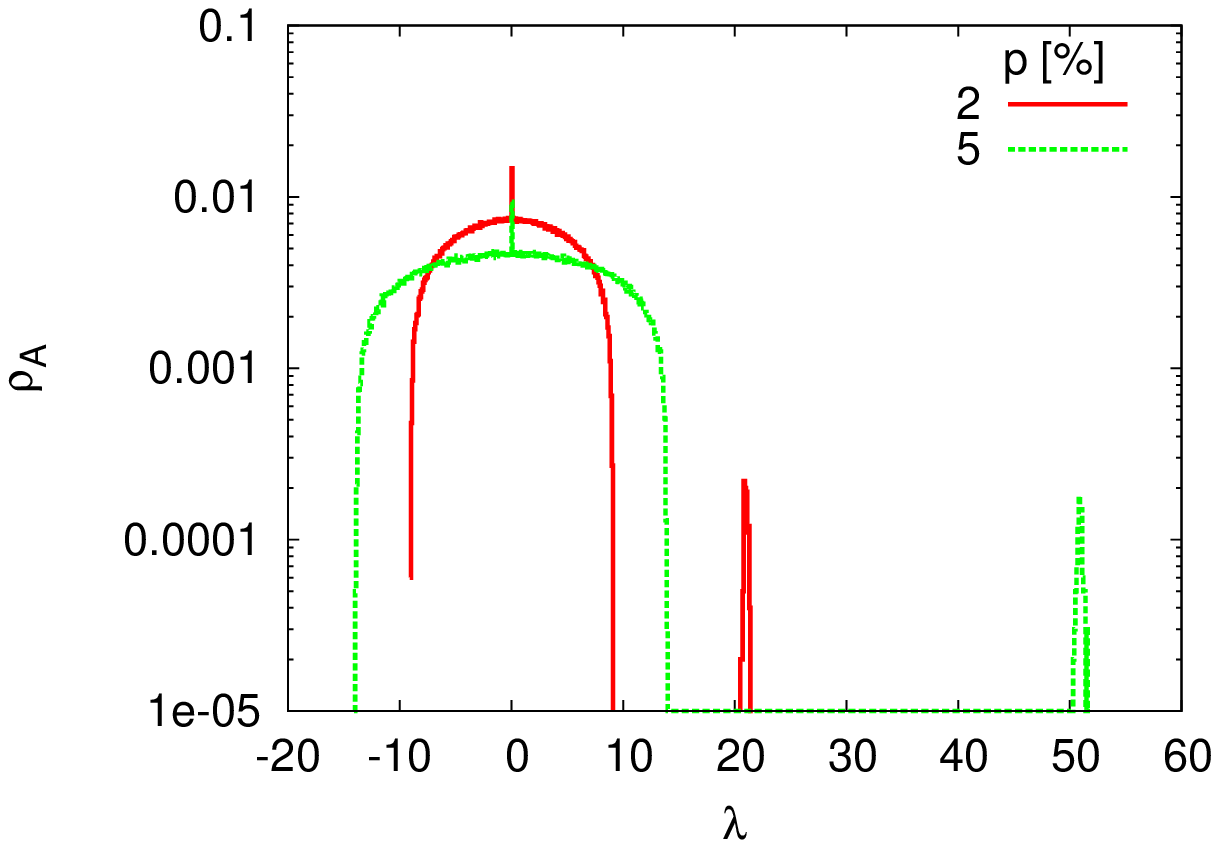}\\
(b) \includegraphics[width=.6\textwidth]{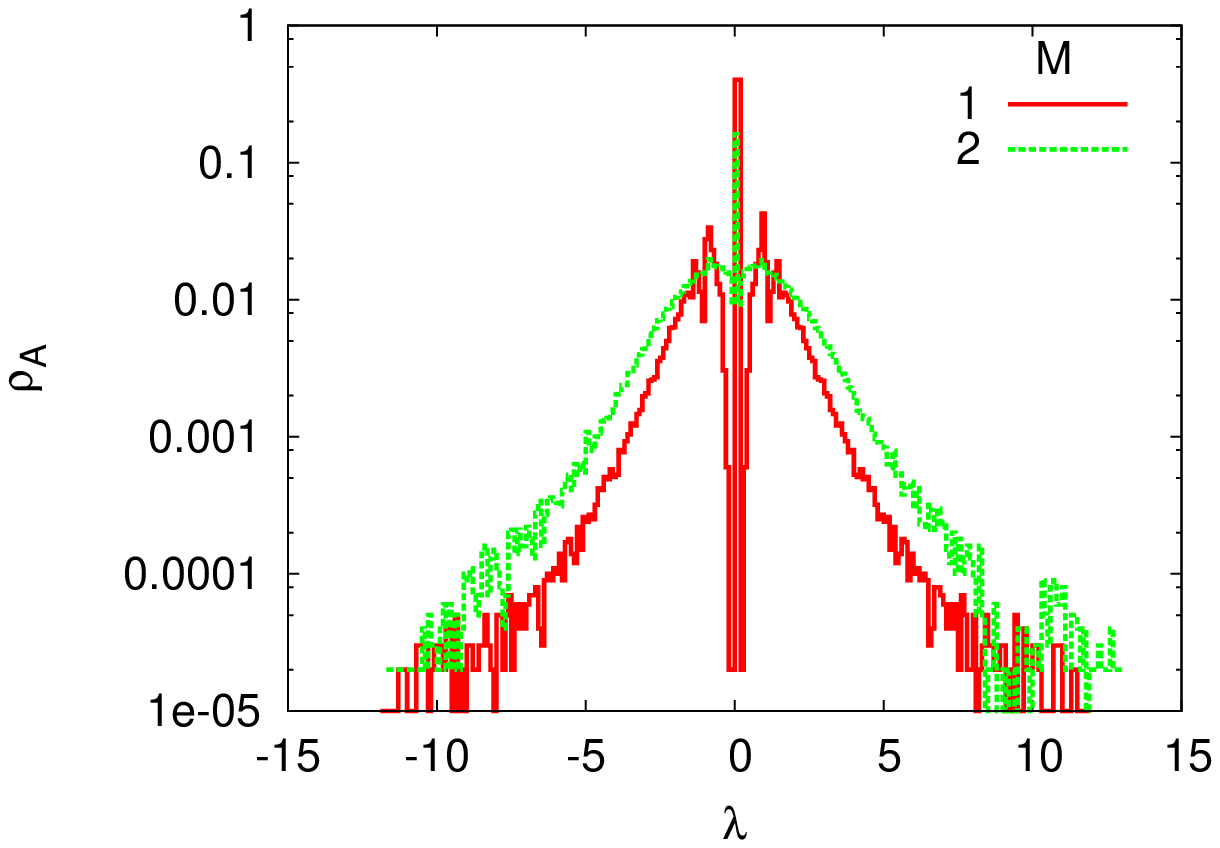}\\
(c) \includegraphics[width=.6\textwidth]{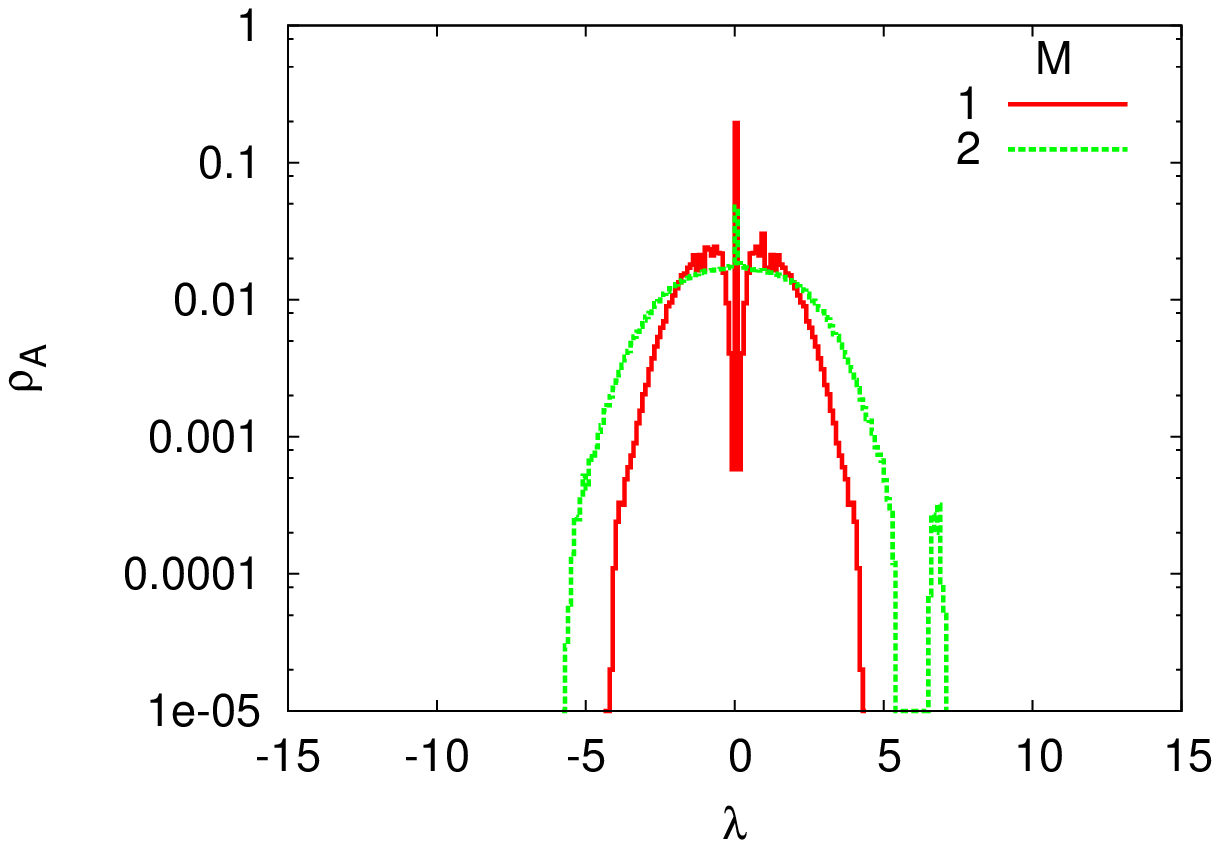}
\caption{Density of eigenvalues $\rho_A(\lambda)$ for adjacency matrices $\mathbf{A}$ for (a) ER, (b) AB and (c) exponential networks with $N=10^3$.
The results are averaged over $N_{\text{run}}=100$ simulations and binned ($\Delta\lambda=0.1$).
The isolated peaks in Fig. \ref{fig-A}(a) correspond to the principal eigenvalue.}
\label{fig-A}
\end{center}
\end{figure}

For the adjacency matrix of ER, the density of eigenvalues
 consist two separated parts: the Wigner-semicircle centered over $\lambda=0$ and with 
radius approximately equal to $2\sqrt{Np(1-p)}$, and the single Frobenius--Perron principal 
eigenvalue near $Np$ \cite{Mehta,Cvetkovic,eigen} (see Fig. \ref{fig-A}(a)).

The detailed study of graph spectrum for AB graphs may be found in Ref. \cite{Farkas-2001,Farkas-2002} by Farkas et al. 
There, the deviation for semicircular law was observed and $\rho_A(\lambda)$ has
 triangle-like shape with power law decay \cite{Farkas-2001}.
A very similar situation occurs for the exponential networks, but $\rho_A(\lambda)$ 
at the top of the ``triangle'' is now more rounded.
The separated eigenvalues are not observed for this kind of networks (see Fig. \ref{fig-A}(b-c)).

Let us discuss the spectrum of eigenvalues of
adjacency matrices of dense graphs in the ER model \cite{Burda-2007}.
The diagonal elements of these matrices
are equal zero $a(i,i)=0$
while the off-diagonal elements $a(i,j)$
assume the value $1$ with the probability $p$
or $0$ with the probability $1-p$.
The elements $a(i,j)$ above
the diagonal are independent identically distributed
random numbers with the probability distribution 
$P(a(i,j)) = (1-p) \delta(a(i,j)) +  p \delta(1 - a(i,j))$.
This probability distribution of $a(i,j)\equiv x$ has the mean value:
$x_0= \langle x \rangle = p$ and the variance 
$\sigma^2 = \langle x^2 \rangle - \langle x \rangle^2 = p(1-p)$.
The universality tells us that the spectrum of random
matrices does not depend on the details of the probability
distribution but only on its mean value and variance:\footnote{if the
variance is finite} the eigenvalue spectrum in the limit 
$N\rightarrow \infty$ is identical for different distributions
as long as they have the same mean and variance. 
In particular one can take a Gaussian
distribution: $1/\sqrt{2\pi\sigma^2} \exp \left[- (x-x_0)^2/2\sigma^2 \right]$.
Thus one can expect that the spectrum of adjacency
matrices of ER graphs can be approximated
for large $N$ by the spectrum of matrices with
continuous random variables 
which have the following probability distribution:
\begin{equation}
\begin{split}
\prod_{i} \frac{da(i,i)}{\sqrt{2\pi}} 
\exp\left[ - \frac{a(i,i)^2}{2\sigma^2}\right]
\cdot\prod_{i<j} \frac{da(i,j)}{\sqrt{2\pi}} 
\exp\left[ - \frac{(a(i,j)-p)^2}{2\sigma^2}\right].
\end{split}
\end{equation}
For the diagonal elements the distribution has the mean equal zero
to reflect the fact that the corresponding adjacency
matrix elements $a(i,i)=0$. The last
expression can be written in a compact form:
\begin{equation}
D {\bf A} \exp \left[ - \frac{1}{2\sigma^2} {\bf tr} ({\bf A} - p {\bf C})^2 \right]
=
D {\bf A} \exp \left[ - \frac{1}{2\sigma^2} {\bf tr} {\bf B}^2 \right],
\end{equation}
where 
\[ D {\bf A} \equiv (\sqrt{2\pi})^{-N(N+1)/2} \prod_{i} da(i,i) \prod_{i<j} da(i,j)\]
 is the standard measure
in the set of symmetric matrices. The matrix $\mathbf{B}$ is obtained
from $\mathbf{A}$ by a shift ${\bf B}  = {\bf A} - p {\bf C}$
where $\mathbf{C}$ has the form:
\begin{equation}
\mathbf{C} =
\begin{pmatrix}
0 & 1 & 1 &  \cdots & 1 & 1 & 1 \\
1 & 0 & 1 &  \cdots & 1 & 1 & 1 \\
1 & 1 & 0 &  \cdots & 1 & 1 & 1 \\
  &   &   &  \ddots &   &   &   \\
1 & 1 & 1 &  \cdots & 0 & 1 & 1 \\
1 & 1 & 1 &  \cdots & 1 & 0 & 1 \\ 
1 & 1 & 1 &  \cdots & 1 & 1 & 0 
\end{pmatrix}.
\end{equation}
The spectrum of the matrix $\mathbf{B}$ is given by the
Wigner semi-circle law \cite{Wigner}:
\begin{equation}
\rho_B(\lambda) = \frac{1}{2\pi N\sigma^2} \sqrt{4N\sigma^2 - \lambda^2}.
\label{bulk}
\end{equation}
It has a support $[-2\sqrt{N}\sigma,2\sqrt{N}\sigma]$, where 
$\sigma= \sqrt{p(1-p)}$ as we calculate above.
We want to determine the spectrum of ${\bf A}$.
It is a sum ${\bf A} = {\bf B} + p{\bf C}$
of matrix ${\bf B}$ for which we already 
know the spectrum \eqref{bulk}
and of matrix $p{\bf C}$ whose spectrum consists of an
$(N-1)$-degenerated eigenvalue $-p$ and one 
eigenvalue $p(N-1)$. The low eigenvalue $-p$
mixes with the eigenvalues of $\bf{B}$ leaving the
bulk of the distribution \eqref{bulk} intact while
the eigenvalue $p(N-1)$ adds to
the distribution a well separated peak in the position 
$p(N-1)$ far beyond the support
$\left[-2\sqrt{Np(1-p)},2\sqrt{Np(1-p)}\right]$ of the main part of the 
distribution:
\begin{equation}
\rho_A(\lambda) \approx  
\rho_B(\lambda) + \frac{1}{N} \delta(\lambda - p(N-1)).
\end{equation}
The considerations hold as long as $p$ is finite.
For sparse graphs $p \sim 1/N \to 0$ 
one sees modifications to the presented picture \cite{Burda-2007}.

We note, that the matrix ${\bf C}$ is both the adjacency and distance matrix for a complete graph.
Thus two very sharp peaks at $\lambda=-1$ and $\lambda=N-1$ constitute a complete graph spectrum
\[ \rho_C(\lambda)=\frac{N-1}{N} \delta(\lambda+1) + \frac{1}{N}\delta(\lambda-(N-1)). \]

\subsection{Spectral Properties of Distance Matrix}

\begin{figure}
\begin{center}
(a) \includegraphics[width=.6\textwidth]{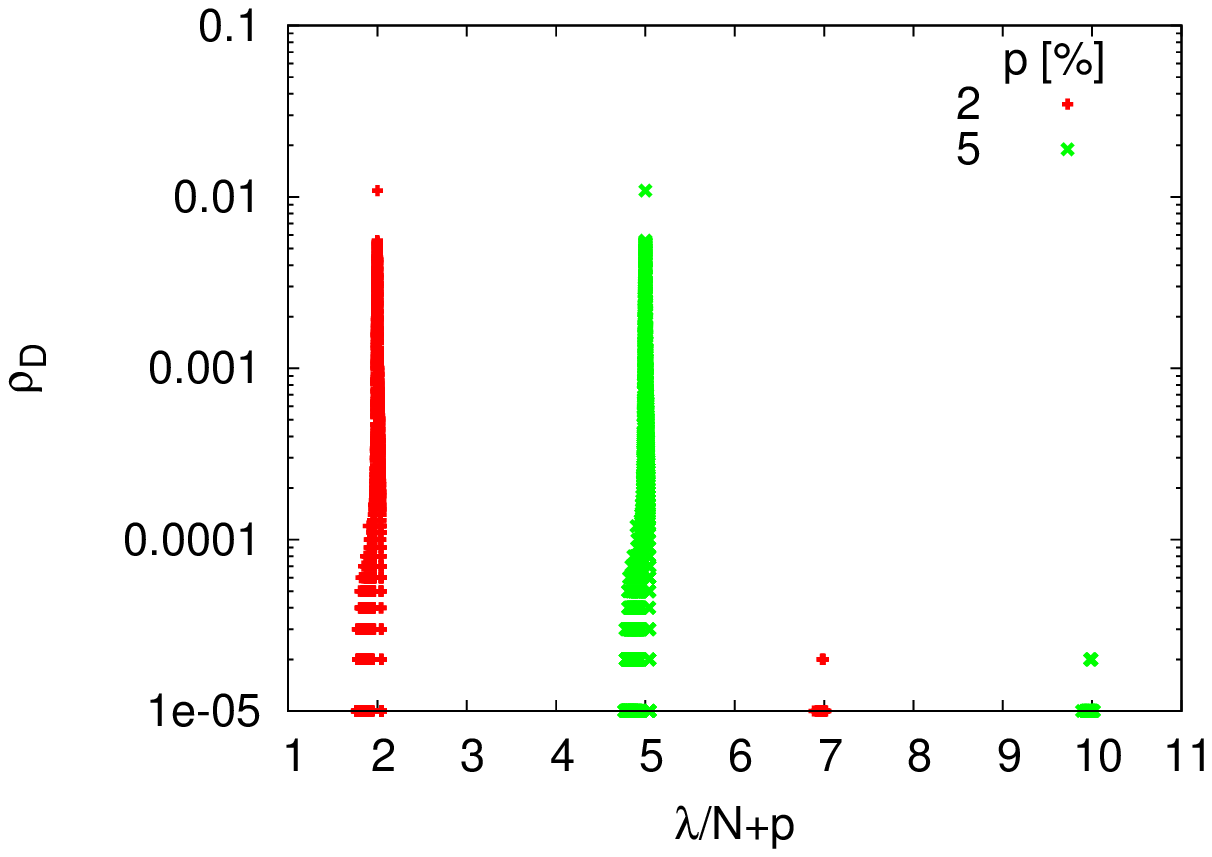}\\
(b) \includegraphics[width=.6\textwidth]{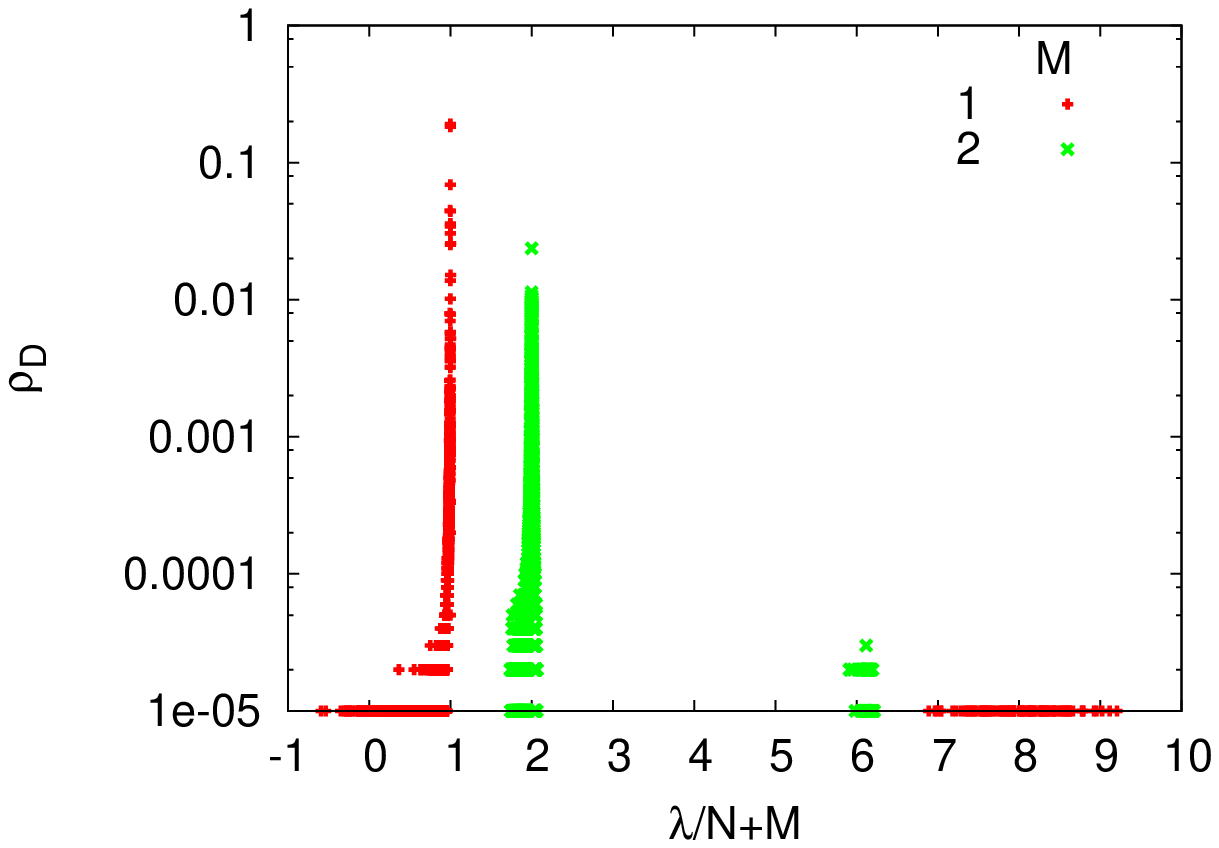}\\
(c) \includegraphics[width=.6\textwidth]{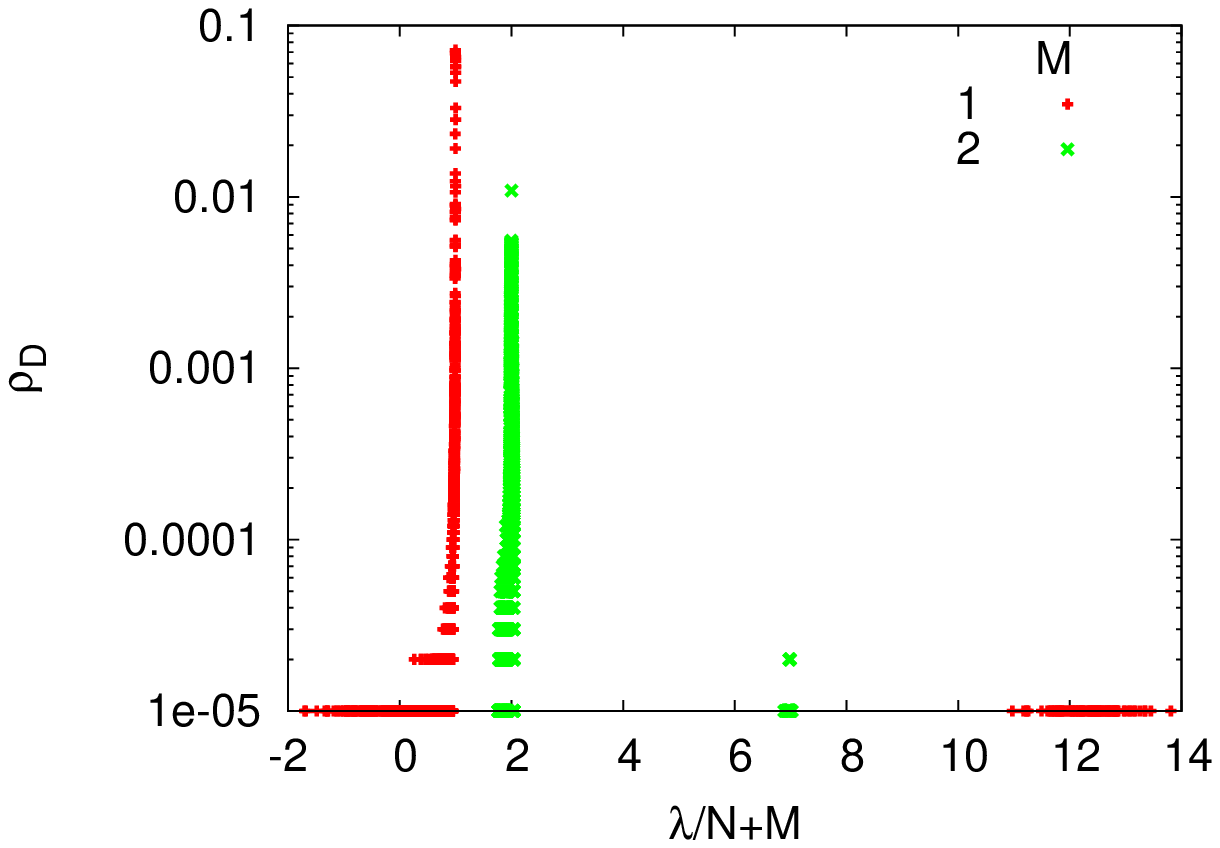}
\caption{Density of eigenvalues $\rho_D$ for distance matrices $\mathbf{D}$ for (a) ER, (b) AB and (c) exponential networks with $N=10^3$.
The results are averaged over $N_{\text{run}}=100$ simulations and binned ($\Delta\lambda=0.1$).
The graphs are horizontally shifted by $M$ or $p$ for better view.}
\label{fig-D}
\end{center}
\end{figure}

\begin{figure}
\begin{center}
\includegraphics[width=.6\textwidth]{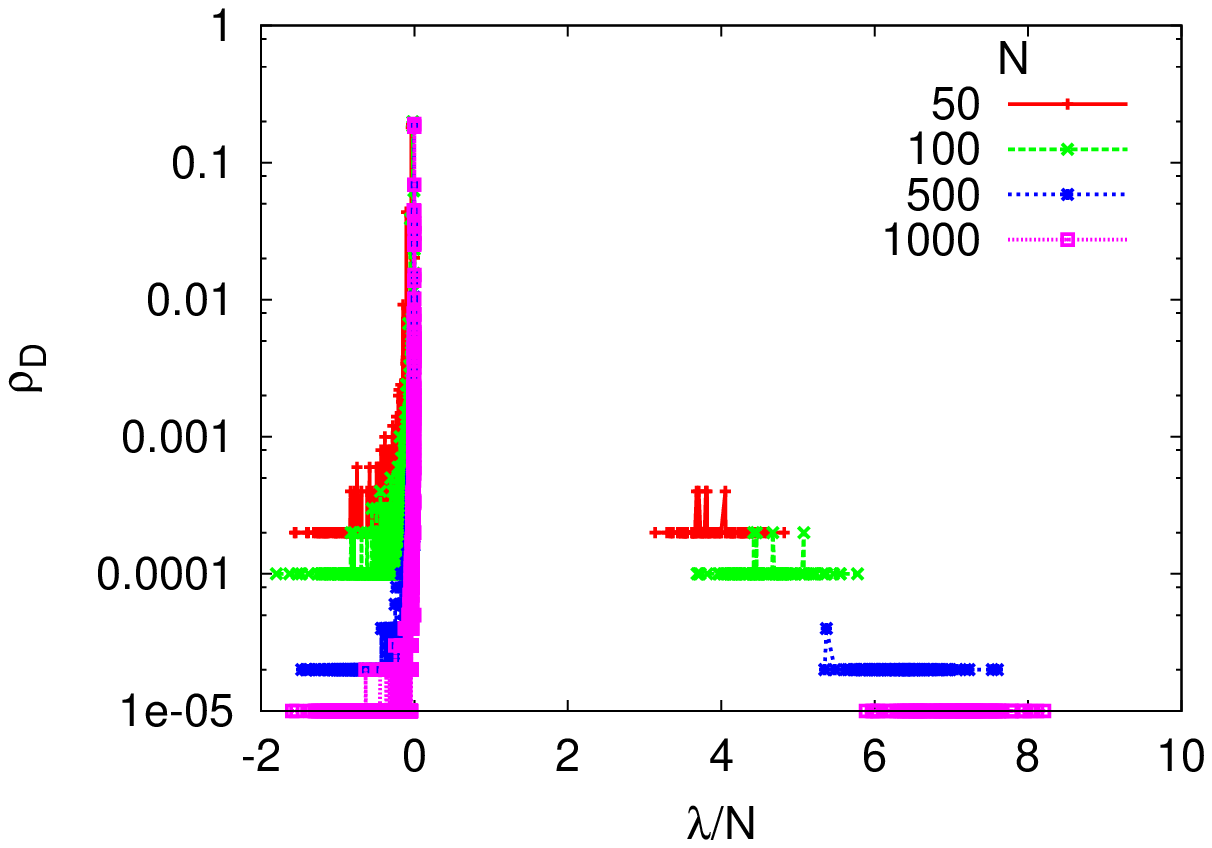}
\caption{Density of eigenvalues $\rho_D(\lambda)$ for distance matrices $\mathbf{D}$ for AB trees with various network size $N$.}
\label{fig-gap}
\end{center}
\end{figure}

Spectra of the distance matrix $\rho_D(\lambda)$ of growing networks for trees ($M=1$) and other graphs ($M>1$) are quantitatively different.
For trees the part of spectrum for $\lambda>0$  is wide and flat.
Moreover, the positive and negative eigenvalues are well separated by a wide gap [see Fig. \ref{fig-D}(b-c)] which increases with networks size $N$ as presented in Fig. \ref{fig-gap}.
On the other hand, we do not observe any finite size effect for negative part of the spectrum.

The density of negative eigenvalues of $\mathbf{D}$ [see Fig.~\ref{fig-D}] is very similar for considered networks.
The positive value part of the spectrum for growing networks does not depend on growth rule and it is roughly the same for AB and exponential networks.

For complete graph ${\bf D}={\bf A}={\bf C}$ and graph spectra consist two sharp peaks as mentioned earlier.

\section{Summary}
\label{sec-summary}

In this paper the spectral properties of the adjacency ${\bf A}$ and distance ${\bf D}$ matrices were investigated for various networks.

For ER and AB networks the well known densities of eigenvalues $\rho_A(\lambda)$ were reproduced.
For the growing networks with attachment kernel $P(k)\propto\text{const}(k)$ the graph spectra are similar to the AB networks except of the spectra center.
For the complete graph two well separated peaks constitute the graph spectrum.

The spectra of distance matrix ${\bf D}$ differ quantitatively for trees and other graphs.
In case of trees ($M=1$) the density of positive eigenvalues is very well separated from the part of the spectrum for $\lambda<0$ and extremely flat.
Thus the specific shape of the distance matrix spectrum may be a signature of absence of loops and cyclic paths in the network.

\subsubsection*{Acknowledgments}
Author is grateful to Zdzis{\l}aw Burda for valuable scientific discussion and to Krzysztof Ku{\l}akowski for critical reading the manuscript.
Part of calculations was carried out in ACK-CYFRONET-AGH.
The machine time on HP Integrity Superdome is financed by the Polish Ministry of Science and Information Technology under Grant No. MNiI/\-HP\_I\_SD/AGH/\-047/2004.


\end{document}